# Multiplicity of Massive Stars at Low Metallicity: Early Results from the BLOeM Campaign

Tomer Shenar[1]
Julia Bodensteiner[2]
Michael Abdul-Masih[3,4]
Frank Backs[5]
Sara R. Berlanas[3,4,6]
Joachim M. Bestenlehner[7,8]
Alexey Bobrick[9,10]
Dominic M. Bowman[5,11]
Nikolay Britavskiy[12]
Paul A. Crowther[8]
Kunal Deshmukh[5]
Matthias Fabry[13]
Mark Gieles[14,15]
Avishai Gilkis[16]
Gemma González-Torà[17]
Götz Gräfener[18]
Maude Gull[19, 20]
Ylva Götberg[21]
Calum Hawcroft[22]
Vincent Hénault-Brunet[23]
Artemio Herrero[3,4]
Gonzalo Holgado[3,4]
Robert G. Izzard[24]
Soetkin Janssens[25]
Harim Jin[26]
Venu M. Kalari[27]
Zehava Katabi[1]
Alex de Koter[2,5]
Jakub Klencki[26]
Brankica Kubátová[28]
Jiří Kubát[28]
Norbert Langer[18]
Thibault Lechien[26]
Daniel Lennon[4,5]
Roel Lefever[17]
Bethany Ludwig[5]
Laurent Mahy[12]
Jesús Maíz Apellániz[6]
Ilya Mandel[9,10]
Armin Mang Román[29]
Grigoris Maravelias[30,31]
Pablo Marchant[32]
Athira Menon[33]
Francisco Najarro[34]
Anna O'Grady[35]
Lidia Oskinova[29]
Roey Ovadia[1]
Lee R. Patrick[34]
Daniel Pauli[5]
Michał Pawlak[36]
Matheus Bernini Peron[17]
Annachiara Picco[5]
Varsha Ramachandran[17]
Mathieu Renzo[37]
Danilo F. Rocha[38]
Hugues Sana[5]
Andreas A.C. Sander[17]
Tom Sayada[1]
Abel Schootemeijer[18]
Fabian R. N. Schneider[17,39]
Rhys Seeburger[40]
Koushik Sen[37]
Sahar Shahaf[41]
Sergio Simón-Díaz[4,5]
Mitchel Stoop[2]
Silvia Toonen[2]
Frank Tramper[34]
Ruggero Valli[26]
Pieterjan Van Daele[11]
Jacco Th. van Loon[42]
Lieke van Son[43]
Alejandro Vigna-Gómez[44]
Jaime I. Villaseñor[41]
Jorick S. Vink[45]
Chen Wang[46,47]
Xiao-Tian Xu[48]

[1] School of Physics and Astronomy, Tel Aviv University, Israel
[2] Anton Pannekoek Institute for Astronomy, University of Amsterdam, Science Park 904, The Netherlands
[3] Canary Islands Institute of Astrophysics, La Laguna, Tenerife, Spain
[4] Department of Astrophysics, University of La Laguna, Tenerife, Spain
[5] Institute of Astronomy, KU Leuven, Belgium
[6] Centre for Astrobiology, Madrid, Spain
[7] School of Chemical, Materials and Biological Engineering, University of Sheffield, UK
[8] School of Mathematical and Physical Sciences, University of Sheffield, UK
[9] School of Physics and Astronomy, Monash University, Clayton, Australia
[10] OzGrav: Australian Research Council Centre of Excellence for Gravitational Wave Discovery, Clayton, Australia
[11] School of Mathematics, Statistics and Physics, Newcastle University, UK
[12] Royal Observatory of Belgium, Brussels, Belgium
[13] Department of Astrophysics and Planetary Science, Villanova University, USA
[14] Pg. Lluís Companys 23, Barcelona, Spain
[15] Institute of Cosmos Sciences (ICCUB), University of Barcelona, Spain
[16] Institute of Astronomy, University of Cambridge, UK
[17] Computational Astronomy Institute, Astronomy Centre, Heidelberg University, Germany
[18] Argelander–Institute for Astronomy, University of Bonn, Germany
[19] Department of Astronomy, California Institute of Technology, Pasadena, USA
[20] The Observatories of the Carnegie Institution for Science, Pasadena, USA
[21] Institute of Science and Technology Austria, Klosterneuburg, Austria
[22] Space Telescope Science Institute, Baltimore, USA
[23] Department of Astronomy and Physics, Saint Mary's University, Halifax, Canada
[24] University of Surrey, United Kingdom
[25] Research Center for the Early Universe, Graduate School of Science, University of Tokyo, Japan
[26] Max Planck Institute for Astrophysics, Garching, Germany
[27] Gemini Observatory/NSF NOIRLab, La Serena, Chile
[28] Astronomical Institute of the Czech Academy of Sciences, Ondřejov, Czech Republic
[29] Institute for Physics and Astronomy, University of Potsdam, Germany
[30] PeriAstron, Heraklion, Crete, Greece
[31] Institute of Astrophysics, FORTH, Heraklion, Greece
[32] Astronomical Observatory, Ghent University, Belgium
[33] Columbia University in the city of New York, Department of Astronomy, U.S.A.
[34] Centre for Astrobiology (CSIC-INTA), Torrejón de Ardoz, Spain
[35] McWilliams Center for Cosmology & Astrophysics, Department of Physics, Carnegie Mellon University, Pittsburgh, USA
[36] Lund Observatory, Division of Astrophysics, Department of Physics, Lund University, Sweden
[37] Steward Observatory, Department of Astronomy, University of Arizona, USA
[38] National Astrophysical Laboratory, Itajubá, Brazil.
[39] Heidelberg Institute for Theoretical Studies, Germany
[40] Astrophysics Research Institute, Liverpool John Moores University, UK
[41] Max Planck Institute for Astronomy, Heidelberg, Germany
[42] Lennard-Jones Laboratories, Keele University, UK
[43] Department of Astrophysics/IMAPP, Radboud University, Nijmegen, The Netherlands
[44] Niels Bohr International Academy, Niels Bohr Institute, University of Copenhagen, Denmark

[45] Armagh Observatory and Planetarium, Northern Ireland, UK
[46] School of Astronomy and Space Science, Nanjing University, People's Republic of China
[47] Key Laboratory of Modern Astronomy and Astrophysics, Nanjing University, People's Republic of China
[48] Tsung-Dao Lee Institute, Shanghai Jiao-Tong University, People's Republic of China

Massive stars at low metallicity ($Z$) play a central role in shaping the high-redshift Universe, yet their multiplicity remains poorly constrained. The Binarity at Low Metallicity (BLOeM) campaign is a two-year survey of 929 stars in the Small Magellanic Cloud with the Fibre Large Array Multi Element Spectrograph (FLAMES) intrument at ESO's Very Large Telescope, providing the first large-scale spectroscopic monitoring of massive stars at low $Z$ ($^1/_5$ *solar*). Analysis of the initial nine epochs reveals high intrinsic binary fractions (> 70%) on the main sequence and a steep decline in evolved objects. Analysis of the full dataset will yield orbital solutions, identify black-hole companions and allow a derivation of the initial mass function for single and binary stars at low $Z$.

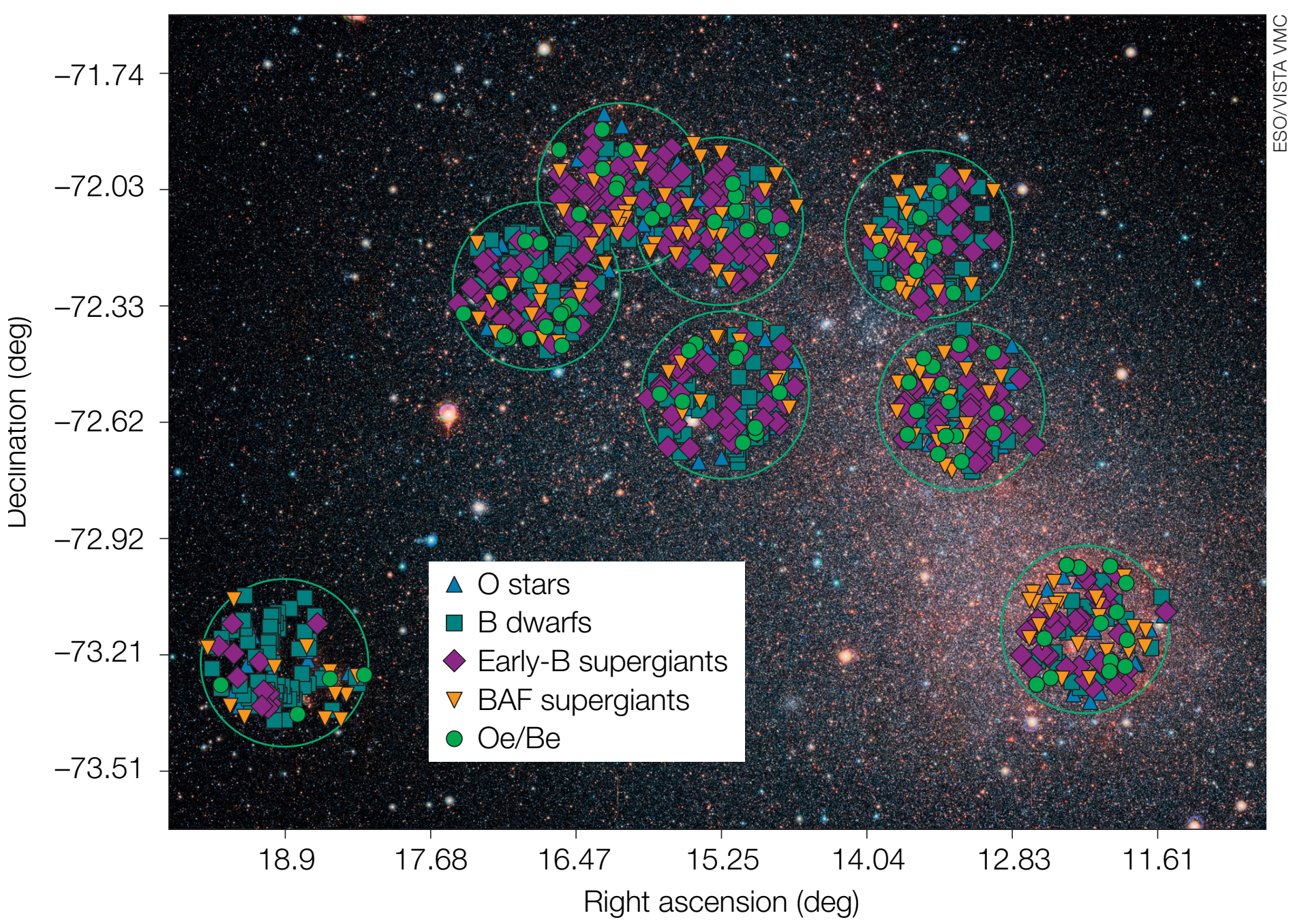

Figure 1. Distribution of the BLOeM targets in the SMC, distributed over eight FLAMES fields (indicated by green circles) and split by spectral types (see legend), overplotted on a VISTA $Y$-$J$-$K_S$ false-colour image.

## Massive-star evolution at low metallicity

Massive stars are born with masses greater than eight times that of our Sun ($M_{initial} \gtrsim 8\ M_{\odot}$). They end their lives in a sudden core collapse into neutron stars and black holes, often associated with a violent supernova explosion. These hot and luminous stars thus serve as a prominent source for mechanical, chemical and radiative feedback in their host galaxies. Surveys of the Large Magellanic Cloud (LMC) and our Milky Way galaxy have revealed unambiguously high multiplicity fractions. In particular, the majority of massive stars interact with a companion star before core collapse (for example, Sana et al., 2012; de Mink et al., 2014). Such binary interactions not only shape the properties of massive-star populations, but also produce unique phenomena such as stripped stars, X-ray binaries and gravitational-wave sources (for example, Langer, 2012; Marchant & Bodensteiner, 2024).

Massive stars with low metallicity ($Z$) are of particular interest as representatives of stellar populations in the early Universe. Their weaker winds are thought to lead to higher rotation rates and more massive stellar remnants. Such stars are predicted to dominate the progenitor population of black-hole and neutron-star mergers detected by gravitational-wave observatories, to produce long gamma-ray bursts and fast radio bursts and to shape the appearance of high-redshift galaxies through intense ultraviolet radiation (Bromm & Yoshida, 2011; Hopkins et al., 2014; Klessen & Glover, 2023). They likely played a role in reionising the early Universe and possibly in the formation of supermassive black holes. Yet our understanding of the multiplicity properties of massive stars and their evolution at low $Z$ remains limited: How do binary fractions and orbital configurations vary with $Z$? Are evolutionary pathways and binary interactions fundamentally different at low $Z$? Does black-hole formation proceed differently?

The Small Magellanic Cloud (SMC; $Z \sim 20\%$ solar metallicity $Z_{\odot}$), hosting about 10 000 massive stars at a distance of 62 kpc, provides the nearest laboratory in which to obtain answers to these questions. The Binarity at Low Metallicity (BLOeM) campaign (Shenar et al., 2024) is an ESO Large Programme (PI: Shenar; dPI: Bodensteiner, programme ID: 112.25R7) designed to characterise the multiplicity of massive stars in the SMC through an unprecedented spectroscopic monitoring campaign with ESO's Very Large Telescope (VLT). BLOeM collected 25 epochs of spectroscopy with the Fibre Large Array Multi Element Spectrograph (FLAMES/GIRAFFE) between October 2023 and September 2025. By measuring radial velocities (RVs) across 25 epochs for two years, BLOeM enables the derivation of binary fractions and orbits, the detailed characterisation of stellar properties, the identification of dormant (i.e., X-ray-quiet) black hole companions, and the inference of the initial mass function of massive stars. While not pristine, the metal content of the SMC is the lowest at which a survey on this scale has been conducted to date.

This article summarises early results from the first nine epochs (Oct–Dec 2023), characterising the stellar content of the survey and providing the first systematic view of massive-star multiplicity across the SMC.

## The BLOeM sample and survey

The 929 BLOeM targets were selected via an automated procedure from the Gaia Data Release 3 catalogue (DR3; Gaia Collaboration et al., 2023) within a radius of 2.6 degrees around the SMC centre, applying magnitude ($G$ < 16.5 mag), colour ($BP$–$RP$ < 1 mag), parallax and proper-motion cuts, and excluding Wolf–Rayet stars and red supergiants (see Shenar et al., 2024 for details). No other prior knowledge about the stars was assumed. The resulting sample is distributed across eight FLAMES fields, each 25 arcminutes in diameter, covering the main star-forming regions of the SMC (Figure 1).

All stars were observed with the FLAMES/GIRAFFE LR02 setting (3950–4570 Å; $R$ ~ 6200), achieving signal-to-noise ≥ 20 (typically 70–100). The 25 epochs, which were scheduled for observing between October 2023 and September 2025, probe orbital periods up to ~1000 days, with the first nine epochs collected between October and December 2023. These nine epochs, which provide a 2–3-month baseline, were co-added to create high-signal-to-noise templates, enabling detailed spectral classification (Shenar et al., 2024).

The sample spans spectral types O4 to F5, with estimated initial masses from ~8 to 80 $M_\odot$. Among the OB stars, 82 exhibit Balmer emission, mostly attributed to rapidly rotating stars with decretion discs (OBe stars). Otherwise, it includes 139 O stars, 309 early B dwarfs/giants (B0-B3 V-III), 262 early B bright giants/supergiants (B0-B3 II-I), and 136 BAF supergiants (B5-F5 I). The sample also contains three B[e] supergiants, four X-ray binaries, three candidate magnetic stars (Of?p), two stripped-star candidates and around 100 eclipsing binaries. Bestenlehner et al. (2025) performed an initial quantitative analysis of the spectra (not yet accounting for binarity), enabling placement on the Hertzsprung–Russell Diagram (HRD), which is shown in Figure 2.

Because massive-star multiplicity is dominated by short-period systems (periods of days to months), the nine-epoch dataset already provides strong leverage on binary fractions, which are described below. However, identifying long-period, low-inclination and extreme mass-ratio binaries will only be possible with the full dataset, as will the complete orbital analysis and characterisation of companions.

## Multiplicity of massive stars at low metallicity

The five subsamples described above probe distinct mass ranges and/or evolutionary phases and their distinct spectral characteristics require tailored measurement techniques. Their multiplicity analysis was therefore done separately. Stars exhibiting significant peak-to-peak RV variations above a chosen threshold (typically 20 km s$^{-1}$) were classified as binaries. Below we summarise the first results in evolutionary order from main-sequence OB stars and OBe stars to evolved supergiants.

### Multiplicity across the main sequence: high also at low metallicity

The O-type stars and the early B-type dwarfs and giants together represent the core hydrogen-burning massive-star population in BLOeM, spanning initial masses from ~15 to 80 $M_\odot$ for O stars down to ~8 to 15 $M_\odot$ for early B-type dwarfs and giants. O-type stars dominate the ionising-photon output and are likely progenitors of the stellar-mass black holes, while early B-type stars are far more numerous, are likely progenitors of neutron stars and trace the bulk of massive-star formation by number. Their combined analysis therefore offers a comprehensive view of massive-star multiplicity across almost an order of magnitude in mass.

Sana, Shenar, & Bodensteiner et al. (2025) identified 62 binaries among 139 O-type stars, yielding an observed fraction of $f_{obs}$ = 45 ± 4%. Accounting for observational biases, they derived an intrinsic (present-day) binary fraction of $f_{int} = 70^{+11}_{-6}$%. This confirms, for the first time, that most O stars in the SMC reside in close binaries, consistent with Galactic and LMC results.

Villaseñor et al. (2025) conducted an equivalent analysis for the 309 early-B dwarfs and giants, identifying 153 binaries among them, corresponding to $f_{obs}$ = 50 ± 3% (see also Moe & Di Stefano, 2013). Bias correction yielded the remarkably high intrinsic fraction of $f_{int}$ = 80 ± 8%, the highest reported for B-type stars at any metallicity. This may hint at an increasing binary fraction toward lower metallicity, though it is consistent with Galactic and LMC results, within errors.

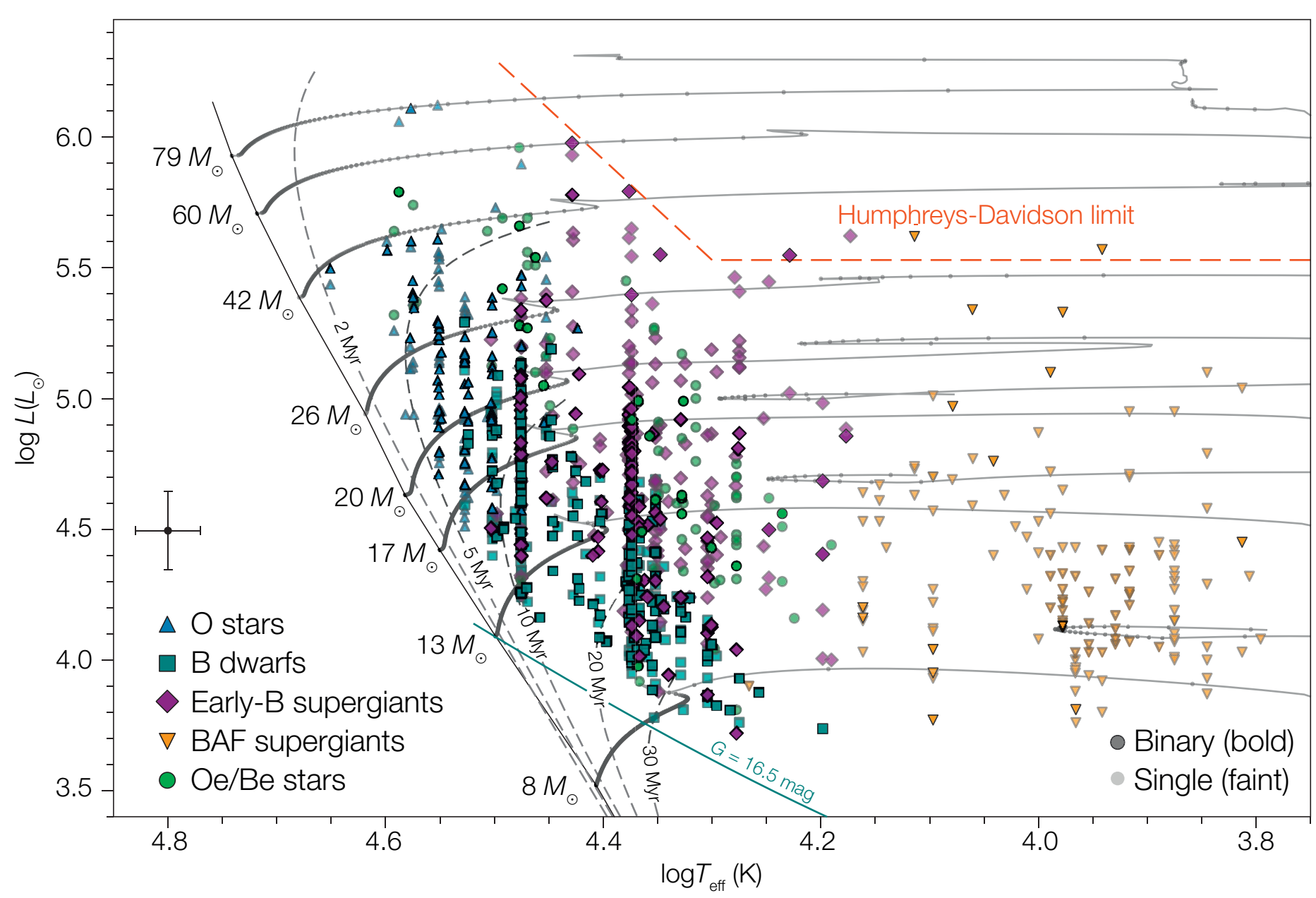

Figure 2. The BLOeM sample shown on a Hertzsprung-Russell diagram (adapted from Shenar et al., 2024), with parameters adopted from Bestenlehner et al. (2025) and Patrick et al. (2025). The assumptions underlying the evolutionary tracks are described by Shenar et al. (2024); dots along the tracks show equidistant time steps of 0.05 Myr. Objects classified as binary candidates from the nine-epoch multiplicity analysis are marked in bold. A very high binary fraction is observed in OB systems, which decreases sharply for the evolved stars in the sample.

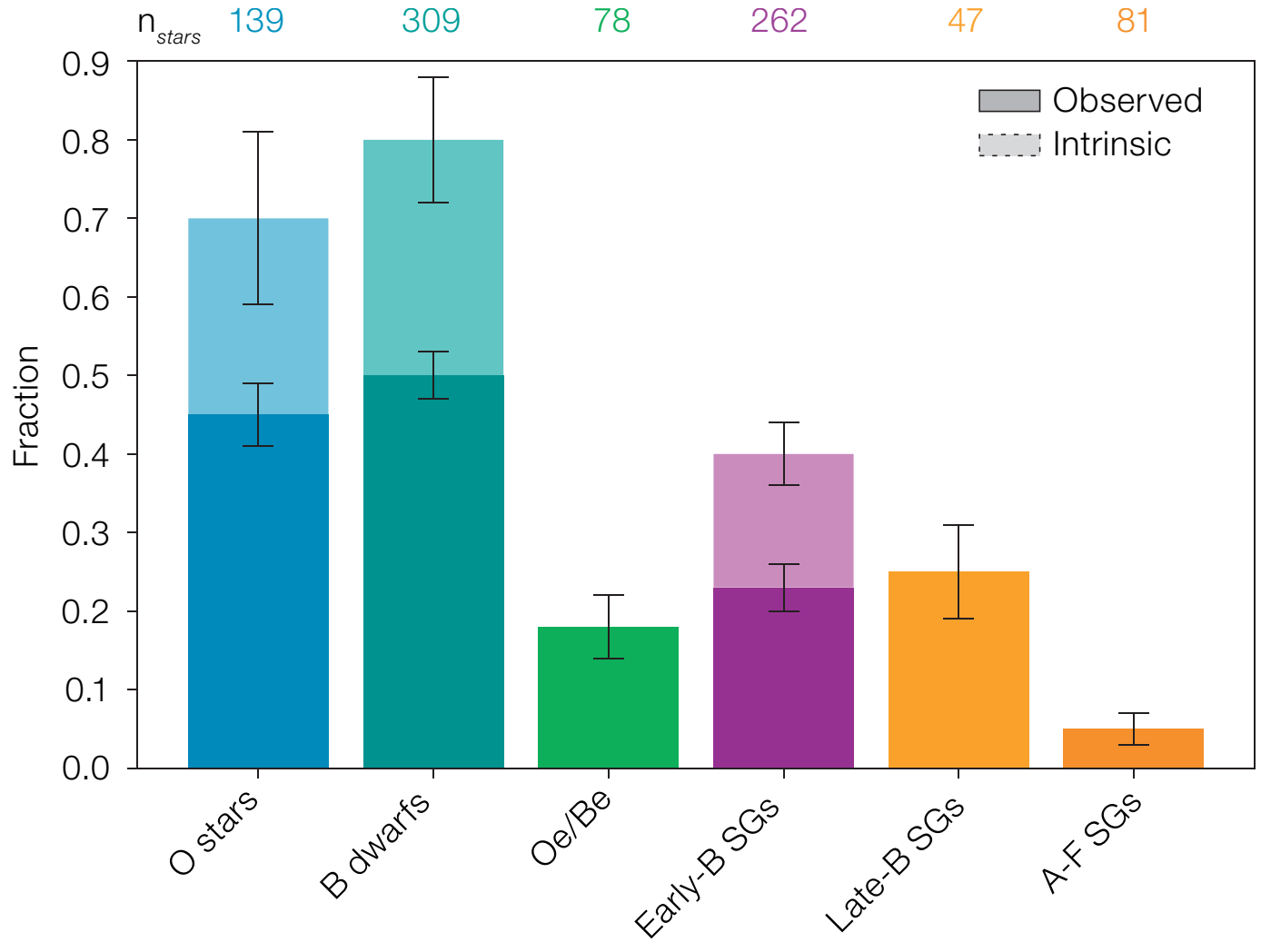

Figure 3. Overview of the binary fractions measured for the different subsamples probed by BLOeM. Results for O stars, B dwarfs, OBe stars, early B supergiants and late BAF supergiants are taken from Sana et al. (2025), Villaseñor et al. (2025), Bodensteiner et al. (2025), Britavskiy et al. (2025) and Patrick et al. (2025). We do not provide bias-corrected binary fractions for likely products of binary interactions (OBe & BAF sample) since knowledge of their underlying orbital-parameter distributions is lacking. Total sizes of the samples are noted above the respective bars.

Together, these two subsamples demonstrate that close binarity is common among main-sequence massive stars at 20% of solar metallicity, with intrinsic fractions of 70–80% across the mass range 8–80 $M_{\odot}$. This is the strongest indication so far that such systems are also common in the metal-poor early Universe, where gravitational-wave mergers originate.

### Multiplicity of classical OBe stars: products of past mass accretion?

The OBe subsample contains 82 stars, mostly characterised by rapid rotation and Balmer emission from decretion discs. Mounting evidence suggests they are binary-interaction products that have accreted mass and angular momentum (for example, Bodensteiner, Shenar & Sana, 2020). In that case, most OBe stars are expected to be classified as single, either because their binary system was disrupted in a supernova explosion or because their companions easily evade detection (i.e., being helium stars or compact objects).

Bodensteiner et al. (2025) reported $f_{obs}$ = 18 ± 4%, increasing to $f_{obs}$ = 32 ± 5% when including candidate binaries. No bias correction was attempted, owing to uncertain underlying orbital properties. The relatively low binary fraction is potentially consistent with the interpretation of this sample as previous mass accretors, but intrinsic variability may have contaminated the results. The full dataset of 25 epochs will be essential to obtaining a robust final picture and characterise potential companions to these objects.

### Multiplicity of B-type supergiants: unexpected residents of the Hertzsprung gap

The sample of early B bright giants & supergiants (B0-B3 II-I) comprises 262 stars. Standard single-star models predict few stars in this part of the HRD (Figure 2): the so-called Hertzsprung gap is expected to be traversed rapidly during post-main-sequence expansion. The large number of B supergiants observed in the SMC stands in stark contrast to models and is therefore puzzling.

One possibility is that many are binary-interaction products (for example, mergers, mass accretors, stripped stars), which would lower the expected binary fraction (for example, Podsiadlowski, Joss & Hsu, 1992; Menon et al., 2024). Alternatively, their presence could indicate that the main sequence extends to cooler temperatures at low $Z$ (for example, Vink et al., 2010; de Burgos et al., 2025). Either way, knowledge of their intrinsic binary fraction is key to unlocking the true nature of blue supergiants.

Britavskiy et al. (2025) reported $f_{obs}$ = 34 ± 3% and $f_{int}$ = 40 ± 4% for this sample, lower than for their OB progenitors, but too high for a population dominated solely by merger products. This may indicate that the sample is a mixture of evolved main sequence objects and binary-interaction products. The full 25-epoch dataset is essential for firm conclusions.

### Multiplicity of BAF supergiants: blue loop stars or binary-interaction products?

The coolest BLOeM subsample consists of 128 B, A and F supergiants (B5 to F5). Their nature is uncertain: they may be a mixture of stars undergoing a so-called blue loop after reaching the red supergiant phase and the products of various types of binary interactions.

Patrick et al. (2025) analysed the first nine epochs for the BAF sample. The exceptional RV precision (~ 0.1 km $s^{-1}$) reported by the authors makes this sample highly susceptible to binary detection. Adopting the 20 km $s^{-1}$ threshold, no binaries are detected within the sample. Relaxing the threshold to 5 km $s^{-1}$, the observed fractions are 25% and 5% in late-type B and AF supergiants, respectively. As few binary systems are known in this evolutionary phase, to compare with BLOeM results Patrick et al. (2025) assumed an underlying orbital parameter distribution typical of main sequence binary systems. The resulting lack of short-period binaries observed within the BAF sample is inconsistent with a gradual evolution from main-sequence stars. These results imply that most BAF supergiants are products of binary interaction. The 25 epochs will be needed to probe longer periods or binaries with extreme mass ratios (for example, low-mass companions or neutron-star companions).

## Towards a comprehensive picture of binarity at low metallicity

Figure 3 provides an overview of the results emerging from analysis of the first nine epochs. The most important BLOeM results with regard to multiplicity are:

– Close binaries are ubiquitous at low metallicity, with intrinsic fractions > 70% in OB-type systems.
– Stellar-evolution models require recalibration: a high density of stars in the Hertzsprung gap and a moderately high binary incidence among them are observed. This implies that the main sequence may extend to cooler temperatures than predicted by standard models, or that the gap is heavily contaminated by binary-interaction products — or both. A dedicated analysis of these results is underway.
– Most BAF supergiants and OBe stars are binary-interaction products, suggested by their drastically lower present-day binary fractions.

## Unleashing the power of BLOeM in upcoming years

The first nine epochs already provide a transformative view of massive-star multiplicity in the SMC. With the full 25 epochs now obtained, BLOeM will soon deliver:
– A deeper binary identification sensitive to long-period (up to about 1000 days) and extreme mass-ratio systems (for example, low-mass companions and neutron-star companions).
– Orbital solutions for hundreds of massive binaries, yielding mass ratios and dynamical masses and including about 100 eclipsing systems.
– Characterisation of companions, via dynamical constraints and techniques such as spectral disentangling.
– A first sample of dormant black holes at low $Z$, providing constraints on black-hole formation, natal kicks and supernova mass ejections (see Willcox et al., 2025).
– A derivation of the initial mass function for binaries and single stars.

Additional studies are underway of rotational velocities, abundance patterns, detailed stellar parameters, runaway stars and individual exotic systems. In addition, a complementary FLAMES/GIRAFFE survey of the BLOeM targets was recently conducted to secure single-epoch observations of the diagnostic He II 4686 and H$\alpha$ lines, enabling the detection of wind features and additional magnetic stars and OBe stars within the sample (PI: Mahy, ID: 115.28A9).

Together, these efforts will illuminate the origins and properties of stellar-mass black holes and neutron stars, gravitational-wave progenitors, and massive stars in general in conditions approaching those of the early Universe.

### Acknowledgements

TS acknowledges support from the European Research Council (ERC) under the European Union's Horizon 2020 research and innovation program (grant agreement 101164755/METAL) and from the Israel Science Foundation (ISF) under grant number 0603225041.

AACS and VR are supported by the Deutsche Forschungsgemeinschaft (DFG, German Research Foundation) in the form of an Emmy Noether Research Group – Project-ID 445674056 (SA4064/1-1, PI Sander).

S.S-D., G.H., and A.H. acknowledge support from the State Research Agency (AEI) of the Spanish Ministry of Science and Innovation (MICIN) and the European Regional Development Fund, FEDER under grants LOS MÚLTIPLES CANALES DE EVOLUCIÓN TEMPRANA DE LAS ESTRELLAS MASIVAS/PRODUCTOS DE LA INTERACCION DE ESTRELLAS MASIVAS REVELADOS POR GRANDES SONDEOS ESPECTROSCOPICOS, with references PID2021-122397NB-C21 / PID2024-159329NB-C21.

G.H received the support from the "La Caixa" Foundation (ID 100010434) under the fellowship code LCF/BQ/PI23/11970035.

DP acknowledges financial support from the FWO in the form of a junior postdoctoral fellowship No. 1256225N.

AP acknowledges support from the FWO under grant agreement No. 11M8325N (PhD Fellowship), and K209924N, K223124N, K1A4925N (Travel Grants).

DMB gratefully acknowledges UK Research and Innovation (UKRI) in the form of a Frontier Research grant under the UK government's ERC Horizon Europe funding guarantee (SYMPHONY; PI Bowman; grant number: EP/Y031059/1), and a Royal Society University Research Fellowship (PI Bowman; grant number: URF\R1\231631).

AO acknowledges support from the McWilliams postdoctoral fellowship.

LRP acknowledges support by grants PID2019-105552RB-C41, PID2022-137779OB-C41 and PID2022-140483NB-C22 funded by MCIN/AEI/10.13039/501100011033 by "ERDF A way of making Europe".

SJ acknowledges funding from the Postdoctoral International Research Fellowships of Japan Society for the Promotion of Science (Graduate school of Science, Tokyo university). JIV acknowledges support from the European Research Council for the ERC Advanced Grant 101054731. FRNS acknowledges support by the Klaus Tschira Foundation. This work has received funding from the European Research Council (ERC) under the European Union's Horizon 2020 research and innovation programme (Grant agreement No. 945806) and is supported by the Deutsche Forschungsgemeinschaft (DFG, German Research Foundation) under Germany's Excellence Strategy EXC 2181/1-390900948 (the Heidelberg STRUCTURES Excellence Cluster). This research is supported by the Flemish Government under the long-term structural Methusalem funding program, project SOUL: Stellar evolution in full glory, grant METH/24/012 at KU Leuven. It is also supported by funding from the Research Foundation – Flanders (FWO) (grant agreement G0ABL24N) and from the KU Leuven Research Council through grant iBOF/21/084.

MP acknowledges the support of the Royal Physio graphic Society in Lund through the Märta and Erik Holmbergs Endowment.

MG acknowledges PID2024-155720NB-I00, CEX2024-001451-M funded by MCIN/AEI/10.13039/501100011033.

IM acknowledges support from the Australian Research Council (ARC) Centre of Excellence for Gravitational Wave Discovery (OzGrav), through project number CE230100016.

RGI thanks STFC for grant ST/Y002350/1.